\documentclass[12pt,a4paper]{article}
\usepackage{xcolor}
\usepackage{jheppub}
\usepackage{epstopdf}
\usepackage{graphicx}
\usepackage{epsfig}
\usepackage{dcolumn}  
\usepackage{bm}    
\usepackage{amssymb} 
\usepackage{amsmath,bm}
\usepackage{amsfonts}    
\usepackage{slashed}  
\usepackage{youngtab}
\usepackage{tikz}
\usetikzlibrary{arrows,decorations.markings}
\usepackage{pgf}
\usetikzlibrary{arrows}
\usepackage[mathscr]{euscript}
\usepackage{epsfig}
\usepackage{pdfpages}
\usepackage{verbatim}
\usepackage{dsfont}
\usepackage{tensor}

\newdimen\tableauside\tableauside=1.0ex
\newdimen\tableaurule\tableaurule=0.4pt
\newdimen\tableaustep
\def\phantomhrule#1{\hbox{\vbox to0pt{\hrule height\tableaurule width#1\vss}}}
\def\phantomvrule#1{\vbox{\hbox to0pt{\vrule width\tableaurule height#1\hss}}}
\def\sqr{\vbox{%
		\phantomhrule\tableaustep
		\hbox{\phantomvrule\tableaustep\kern\tableaustep\phantomvrule\tableaustep}%
		\hbox{\vbox{\phantomhrule\tableauside}\kern-\tableaurule}}}
\def\squares#1{\hbox{\count0=#1\noindent\loop\sqr
		\advance\count0 by-1 \ifnum\count0>0\repeat}}
\def\tableau#1{\vcenter{\offinterlineskip
		\tableaustep=\tableauside\advance\tableaustep by-\tableaurule
		\kern\normallineskip\hbox
		{\kern\normallineskip\vbox
			{\gettableau#1 0 }%
			\kern\normallineskip\kern\tableaurule}%
		\kern\normallineskip\kern\tableaurule}}
\def\gettableau#1 {\ifnum#1=0\let\next=\null\else
	\squares{#1}\let\next=\gettableau\fi\next}

\allowdisplaybreaks

\newcommand{\be}{ \begin{equation}}
\newcommand{\ee}{\end{equation}}

\def\one{{\hbox{ 1\kern-.8mm l}}}
\def\zero{{\hbox{ 0\kern-1.5mm 0}}}

\title{Higher spin algebras and large $\mathcal{N}=4$ holography}

\author{Lorenz Eberhardt$^{a,b}$, Matthias R.\ Gaberdiel$^{a,c}$ and Ingo Rienacker$^a$} 
\affiliation{$^a$ Institut f\"ur Theoretische Physik, ETH Zurich, \\
\hspace*{0.3cm}CH-8093 Z\"urich, Switzerland}
\affiliation{$^b$ School of Natural Sciences, Institute for Advanced Study, \\
\hspace*{0.3cm}Princeton, NJ 08540, USA}
\affiliation{$^c$ Beijing International Center for Mathematical Research \\
\hspace*{0.3cm}Peking University, Beijing 100871, P.R.\ China}
\emailAdd{eberhardtl@itp.phys.ethz.ch, gaberdiel@itp.phys.ethz.ch,  ringo@student.ethz.ch}

\abstract{A new family of higher spin algebras that arises upon restricting matrix extensions of $\mathfrak{shs}[\lambda]$ is found. We identify coset CFTs realising these symmetry algebras, and thus propose new higher spin--CFT dual pairs. These higher spin theories arise naturally as a subsector of string theory on AdS$_3\times {\rm S}^3 \times {\rm S}^3 \times {\rm S}^1$ for specific ratios of the radii of the two spheres.
}

\begin{document}

\maketitle

\makeatletter
\g@addto@macro\bfseries{\boldmath}
\makeatother
\section{Introduction}

Recently, important progress has been made in elucidating the duality between higher spin theories on AdS$_3$ \cite{Prokushkin:1998bq,Prokushkin:1998vn} and $2d$ conformal field theories \cite{Henneaux:2010xg,Campoleoni:2010zq,Gaberdiel:2010pz}, see \cite{Gaberdiel:2012uj} for a review.  The relevant higher spin algebras that appear in these examples all arise from the prototypical higher spin algebra $\mathfrak{shs}[\lambda]$  by means of a number of small constructions \cite{Vasiliev:1999ba}. In particular, one may either extend the higher spin algebra by adding Chan-Paton (or matrix) degrees of freedom, or one may restrict these algebras by imposing some invariance conditions that reduce, say the `unitary' algebra $\mathfrak{shs}[\lambda]$ to its `orthogonal' or `symplectic' version.

It was shown in \cite{Gaberdiel:2014cha} that the $\lambda=0$ point of the ${\cal N}=4$ version of the duality embeds naturally into the stringy AdS/CFT duality. The point $\lambda=0$ is special since it corresponds to the geometry AdS$_3\times {\rm S}^3 \times \mathbb{T}^4$ --- for general $\lambda$ the geometry is expected to be AdS$_3\times {\rm S}^3 \times {\rm S}^3 \times {\rm S}^1$ --- and since the CFT dual of string theory on AdS$_3\times {\rm S}^3 \times \mathbb{T}^4$ has long been known: it is the symmetric orbifold of $\mathbb{T}^4$, see e.g.~\cite{David:2002wn} for a review. More specifically, it was shown in \cite{Gaberdiel:2014cha} that the ${\cal N}=4$ superconformal ${\cal W}_{\infty}[0]$ symmetry of the CFT dual of the higher spin theory is a natural subsector of the symmetric orbifold of $\mathbb{T}^4$. This reflects nicely the general expectation about how higher spin theories should be related to string theories. 

At the time of \cite{Gaberdiel:2014cha}, the CFT dual of string theory for AdS$_3\times {\rm S}^3 \times {\rm S}^3 \times {\rm S}^1$ was not known, but recently convincing evidence for a specific dual CFT was given in \cite{Eberhardt:2017pty}. The proposal of \cite{Eberhardt:2017pty} only covers the cases when the ratio of the D5-brane charges takes the values (here we assume, without loss of generality, that $Q_5^+ \leq Q_5^-$)
\be\label{lambdaspec}
\gamma = \frac{Q_5^+}{Q_5^+ + Q_5^-} = \frac{1}{1+n} \ , \qquad n=1,2,\ldots \ , 
\ee
and even for these values, the dual CFT does not, in general, seem to contain a natural higher spin subalgebra. On the other hand, for $\gamma=\frac{1}{2}$ a certain ${\cal N}=2$ supersymmetric higher spin subalgebra was already identified in \cite{Eberhardt:2017pty}, see also Section~\ref{T2algebra} below. In this paper we analyse another natural construction that generalises more directly to the other (admissible) values of $\gamma$, see the discussion in Section~\ref{sec:otheralpha}. While the relevant higher spin algebras are not supersymmetric, one may suspect that they coincide with standard constructions from \cite{Vasiliev:1999ba}. For example, for the theory at $\gamma=\frac{1}{2}$, the most natural candidate is the so-called $\mathfrak{ho}(1|4)$ algebra of \cite{Vasiliev:1999ba} since it has a matching spin spectrum.

In this paper we show that this expectation is not quite borne out. In particular, the relevant algebra at $\gamma=\frac{1}{2}$ does not agree with $\mathfrak{ho}(1|4)$, but rather involves another class of higher spin algebra constructions that does not seem to have appeared before in the literature. We will therefore explain this construction in some generality. All of the resulting algebras define consistent higher spin theories, and it is not difficult to identify the corresponding dual ${\cal W}_\infty$ algebras, as well as coset realisations of them. Thus these higher spin algebras also fit naturally into the family of higher spin--CFT dual pairs. 
\medskip

The paper is organized as follows. In Section~\ref{sec:general} we explain the general construction of the higher spin algebras, and describe some basic properties of them. In Section~\ref{sec:CFTconstructions} we construct the dual ${\cal W}_\infty$ algebras, and explain how they can be realised by free field constructions (for one limiting value of $\lambda$), as well as by cosets. In Section~\ref{sec:applications} we then show how these theories arise naturally in the context of large ${\cal N}=4$ holography. Our conclusions are summarised in Section~\ref{sec:conclusions}, and there are five appendices into which some of the more technical material has been delegated. 

\section{The general construction}\label{sec:general}

In this section we explain a novel general construction for how to define what one may call a restricted matrix extension of $\mathfrak{hs}[\lambda]$ or $\mathfrak{shs}[\lambda]$. In the following, we shall concentrate on the case of $\mathfrak{shs}[\lambda]$, from which the construction for $\mathfrak{hs}[\lambda]$ can be easily deduced.

\subsection{The higher spin algebra $\mathfrak{shs}[\lambda]$}

Let us begin by recalling the definition of the prototypical higher spin algebra $\mathfrak{shs}[\lambda]$. By definition, $\mathfrak{shs}[\lambda]$ is a Lie algebra, whose Lie bracket is induced from an associative product, often called the `lone star-product', on the space $\mathds{1} \oplus \mathfrak{shs}[\lambda]$, where $\mathds{1}$ denotes the identity. For the following it will be more convenient to work directly with this associative algebra, and we shall, by slight abuse of notation, also denote it by $\mathfrak{shs}[\lambda]$.

We may define $\mathfrak{shs}[\lambda]$ as the quotient of the universal enveloping algebra of $\mathfrak{osp}(1|2)$ by the two-sided ideal generated by $\mathcal{C}-\tfrac{1}{4}\lambda(\lambda-1)\mathds{1}$, where $\mathcal{C}$ is the quadratic Casimir of $\mathfrak{osp}(1|2)$. This algebra has an equivalent description in terms of oscillators $k$ and $y_\alpha$ with $\alpha\in\{1,2\}$, satisfying \cite{Bergshoeff:1990cz, Bergshoeff:1991dz, Prokushkin:1998bq, Prokushkin:1998vn}
\be 
[y_\alpha,y_\beta]=2\mathrm{i}\epsilon_{\alpha\beta}(1+\nu k)\ , \quad k y_\alpha=-y_\alpha k\ , \quad k^2=1\ , \label{defining commutators}
\ee
where $\nu=2\lambda-1$. The algebra is actually ${\cal N}=2$ superconformal and hence contains $\mathfrak{osp}(2|2) \cong \mathfrak{sl}(1|2)$ as a subalgebra, see, e.g.~\cite[Section~2.1]{Gaberdiel:2013vva} for a detailed description. In terms of this $\mathfrak{sl}(1|2)$, the higher spin algebra consists of the following (long) $\mathfrak{sl}(1|2)$ multiplets:
\be 
\mathfrak{shs}[\lambda]\cong\mathds{1}\oplus \mathfrak{sl}(1|2) \oplus \bigoplus_{s=2}^\infty R^{(s)}\ ,
\ee
where the $\mathcal{N}=2$ multiplet $R^{(s)}$ consists of the states 
\begin{align}
R^{(s)} : \begin{tabular}{cc}
$s:$ & $(0)$ \\
$s+\tfrac{1}{2}:$ & $(+) \oplus (-)$   \\
$s+1:$ & $(0)$
\end{tabular}.
\end{align}
Here $s$ (resp.\ $s+\frac{1}{2}$ and $s+1$) denotes the spin of the corresponding spacetime fields, which is related to the spin $j$ with respect to the M\"obius subalgebra as $j=s-1$, while the quantum numbers in brackets refer to the eigenvalue with respect to the $\mathfrak{u}(1)$ current $J_0$, see \cite{Gaberdiel:2013vva} for our conventions. In terms of the oscillators, the spin simply equals $s=\tfrac{m}{2}+1$, where $m$ is the number of $y_\alpha$ generators in the expression. On the other hand, the fermionic generators with $\pm$ $\mathfrak{u}(1)$ charge are proportional to $(1\pm k)$. 

In this construction $\lambda$ (or $\nu$) is a free parameter, and we have the symmetry $\lambda \leftrightarrow 1-\lambda$, as can be seen by exchanging $k$ with $-k$; more formally, this means that we have an algebra isomorphism
\be
\Phi:\, \mathfrak{shs}[\lambda] \longrightarrow \mathfrak{shs}[1-\lambda] \ , \qquad (k,y_\alpha) \longmapsto (-k,y_\alpha)\ . \label{k minus k isomorphism}
\ee

\subsection{The restricted matrix extension}\label{matrix extension}

Now we want to extend this construction to include matrix degrees of freedom.  To this end, let us first consider the associative algebra
\be 
\mathfrak{shs}_M[\lambda] \equiv \mathfrak{shs}[\lambda] \otimes \mathrm{Mat}(M,M,\mathbb{C})\ ,
\ee
where $M$ is a positive integer, and $\mathrm{Mat}(M,M,\mathbb{C})$ is the (associative) algebra of complex $M\times M$ matrices.
This algebra is well-known in the literature; in particular the special case $\mathfrak{shs}_2[\lambda]$ leads to the higher spin algebra with large $\mathcal{N}=4$ supersymmetry \cite{Gaberdiel:2013vva}. The algebra $\mathfrak{shs}_M[\lambda]$ was also considered in \cite{Creutzig:2013tja}.

We now subdivide $M=K+L$, and decompose the $M\times M$ matrices into blocks of the form 
\be
\left( \begin{matrix} A & B \\ C & D \end{matrix} \right) \ , 
\ee
where $A$ is a $K\times K$ matrix, $B$ a $K\times L$ matrix, $C$ a $L\times K$ matrix, and finally $D$ a $L\times L$ matrix. 
Associated with this subdivision, we then define $\mathfrak{shs}(K|L)[\lambda]$ to be the subalgebra of $\mathfrak{shs}_M[\lambda]$ that is generated by the elements of the form
\be 
\begin{pmatrix}
y^{2m}(1-k) \otimes A & 0 \\
0 & y^{2m}(1+k) \otimes D
\end{pmatrix}\text{ and } \begin{pmatrix}
0 & y^{2n+1}(1+k) \otimes B \\
y^{2n+1}(1-k) \otimes C & 0
\end{pmatrix}\ ,\label{hu lambda definition}
\ee
where $m$ and $n$ are nonnegative integers.\footnote{Here $y^l$ stands for any string of $y_\alpha$ of total length $l$.} The first kind of generators have spin $m+1$ and are bosonic, while the second kind have spin $n+\tfrac{3}{2}$ and are fermionic. Using the defining relations \eqref{defining commutators}, one can easily verify that this algebra closes. (In particular, $\tfrac{1}{2}(1+k)$ and $-\tfrac{1}{2}(1-k)$ are projectors, and the parity constraint we imposed on the number of $y$'s is preserved by the matrix multiplication.) 

Alternatively, we may characterize the resulting algebra $\mathfrak{shs}(K|L)[\lambda]$ as the subalgebra of $\mathfrak{shs}_{M}[\lambda]$ on which $k \mathds{1}_{M}$ acts by $\mathrm{diag}(-\mathds{1}_{K},\mathds{1}_{L})$, both from the left and from the right. We should note that this truncation differs from those presented in \cite{Vasiliev:1999ba}, since it mixes the $\mathfrak{shs}[\lambda]$ part with the matrix part in a non-trivial way. 

For the special case of $\lambda=\tfrac{1}{2}$, the generator $k$ can be decoupled from the algebra, as can be seen from the defining relation \eqref{defining commutators}. In this case, our construction degenerates to the one of \cite{Vasiliev:1999ba}. Indeed, in the product of the generators of \eqref{hu lambda definition}, the $(1+ k)$ term never meets the $(1-k)$ term, but we only need the relations  $(1\pm k)^2 = \pm 2 \, (1\pm k)$. Thus, we have an algebra isomorphism
\begin{equation}
	\begin{pmatrix}
		y^{2m}(1-k) \otimes A & 0\\
		0 & y^{2m}(1+k) \otimes D
	\end{pmatrix}
	\mapsto
	\begin{pmatrix}
		-2y^{2m} \otimes A & 0 \\
		0 & 2y^{2m} \otimes D
	\end{pmatrix},
\end{equation}
\begin{equation}
	\begin{pmatrix}
	0 & y^{2n+1}(1+k) \otimes B\\ 
	y^{2n+1}(1-k) \otimes C & 0
	\end{pmatrix}
	\mapsto
	\begin{pmatrix}
	 0 & 2y^{2n+1} \otimes B\\
	 -2y^{2n+1}\otimes C & 0
	\end{pmatrix} \ , 
\end{equation}
hence identifying our algebra with the matrix construction of \cite{Vasiliev:1999ba}.

\subsection{Further truncations}\label{further truncations}

As is familiar from other higher spin constructions, we may further truncate these algebras to the orthogonal and symplectic analogues. In particular, we define the `orthogonal truncation' $\mathfrak{sho}(K|L)[\lambda]$ as the subalgebra consisting of those generators $\xi(y,k)\in \mathfrak{shs}(K|L)[\lambda]$ that are invariant under the anti-automorphism
\be 
\sigma(\xi(y,k))=-\mathrm{i}^{|\xi|}\,\xi(\mathrm{i}y,k)^{\mathrm{T}}\ . \label{so automorphism}
\ee
Here, $y=(y_1,y_2)$ is the collection of $y$'s, while the transpose $\mathrm{T}$ acts on the matrix part only. $|\xi|$ denotes the grading of $\xi(y,k)$, i.e.~the power of $y$ in the expression modulo 2. \eqref{so automorphism} is only defined for elements of pure degree, and the definition is then extended by linearity. 
This truncation yields the following restrictions on the matrices $A$, $B$, $C$ and $D$ of \eqref{hu lambda definition}:
\begin{alignat}{3}
s &\in 2\mathbb{Z}+1: \quad && A = - A^{\mathrm{T}} \ , \quad && D = - D^{\mathrm{T}} \ , \\
s &\in 2\mathbb{Z}: \quad && A =  A^{\mathrm{T}} \ , \quad && D = D^{\mathrm{T}} \ , \\
s &\in 2\mathbb{Z}+\tfrac{1}{2}: \quad && B = C^{\mathrm{T}}\ , \\
s &\in 2\mathbb{Z}+\tfrac{3}{2}: \quad && B = -C^{\mathrm{T}}\ . \label{orthogonal restrictions}
\end{alignat} 

For the `symplectic' truncation we assume that $M$, $K$ and $L$ are all even. Then we define $\mathfrak{shsp}(K|L)[\lambda]$ as the subalgebra consisting of 
\be 
\sigma(\xi(y,k))=-\mathrm{i}^{|\xi|}\,\Omega\,\xi(\mathrm{i}y,k)^{\mathrm{T}}\,\Omega^{-1}\ , \label{sp automorphism}
\ee
where we have used the conventions of Appendix~\ref{sec:orthsymp}. (In particular, $\Omega$ is the symplectic form on $\mathbb{R}^{K+L}$.) This translates into the condition that the block-matrix entries in \eqref{hu lambda definition} satisfy 
\begin{alignat}{3}
s &\in 2\mathbb{Z}+1: \quad && A = - \Omega_K A^{\mathrm{T}}  \Omega_K^{-1} \ , \quad && D = - \Omega_L\, D^{\mathrm{T}} \Omega_L^{-1} \ ,  \\
s &\in 2\mathbb{Z}: \quad && A =  \Omega_K A^{\mathrm{T}}  \Omega_K^{-1} \ , \quad && D =  \Omega_L\, D^{\mathrm{T}} \Omega_L^{-1} \ , \\
s &\in 2\mathbb{Z}+\tfrac{1}{2}: \quad && B = \Omega_K^{-1}C^{\mathrm{T}}\Omega_L\ , \\
s &\in 2\mathbb{Z}+\tfrac{3}{2}: \quad && B = -\Omega_K^{-1}C^{\mathrm{T}}\Omega_L\ .
\end{alignat}

\subsection{Properties}

There are a few basic properties of these algebras that may be worth pointing out. Each of them contains generators of spin $s=1$, and they generate the Lie algebras $\mathfrak{u}(K) \oplus \mathfrak{u}(L)$ for the case of $\mathfrak{shs}(K|L)[\lambda]$, and similarly in the other cases.\footnote{Strictly speaking, the relevant Lie algebras are the corresponding complexifications, e.g.\ $\mathfrak{gl}(K;\mathbb{C}) \oplus \mathfrak{gl}(L;\mathbb{C})$ for the case of $\mathfrak{shs}(K|L)[\lambda]$, but as is common in physics, we will often use the compact real form.} As a consequence, all other generators transform 
in representations of this algebra, and we have summarized the representations that appear in Table~\ref{tab:spin content}. (Our conventions regarding the representations of the orthogonal and symplectic groups are given in Appendix~\ref{sec:orthsymp}.)

\begin{table}[h]
\begin{center}
\begin{tabular}{cccc} \\
spin & $\mathfrak{shs}(K|L)[\lambda]$ \\ 
\hline
$\mathbb{Z}$ & $\mathfrak{u}(K) \oplus \mathfrak{u}(L)$ \\
$\mathbb{Z}+\tfrac{1}{2}$ & $(\mathbf{K},\bar{\mathbf{L}})\oplus (\bar{\mathbf{K}},\mathbf{L})$  \\[5pt]

spin & $\mathfrak{sho}(K|L)[\lambda]$ \\ 
\hline
$2\mathbb{Z}+1$ & $([0,1,\dots], [0,1,\dots])$ \\
$2\mathbb{Z}$ & $([2,0,\dots] \oplus [0,0,\dots],[2,0,\dots] \oplus [0,0,\dots])$ \\
$\mathbb{Z}+\tfrac{1}{2}$ & $([1,0,\dots],[1,0,\dots])$ \\[5pt]

spin &$\mathfrak{shsp}(K|L)[\lambda]$ \\
\hline
$2\mathbb{Z}+1$& $([2,0,\dots],[2,0,\dots])$ \\
$2\mathbb{Z}$ & $([0,1,\dots]\oplus [0,0,\dots],[0,1,\dots]\oplus [0,0,\dots])$\\
$\mathbb{Z}+\tfrac{1}{2}$& $([1,0,\dots],[1,0,\dots])$ \\
\end{tabular}
	
\caption{Spin content of the various higher spin algebras. $\mathbf{K}$ denotes the fundamental (vector) representation. In the case of $\mathfrak{u}(K)$, $\bar{\mathbf{K}}$ denotes the anti-fundamental representation.} \label{tab:spin content}
\end{center}
\end{table}

The automorphism of $\mathfrak{shs}[\lambda]$ described in (\ref{k minus k isomorphism}) implies that we have the dualities 
\begin{align} 
\mathfrak{shs}(K|L)[\lambda] &\cong \mathfrak{shs}(L|K)[1-\lambda]\ , \nonumber\\
\mathfrak{sho}(K|L)[\lambda] &\cong \mathfrak{sho}(L|K)[1-\lambda]\ , \label{duality} \\
 \mathfrak{shsp}(K|L)[\lambda] &\cong \mathfrak{shsp}(L|K)[1-\lambda]\ . \nonumber
\end{align}
In fact, we could have reversed the roles of $k$ and $-k$ in the definition \eqref{hu lambda definition} to obtain also the algebra $\mathfrak{shs}(K|L)[1-\lambda]$ as a subalgebra of $\mathfrak{shs}_{K+L}[\lambda]$. Since $\tfrac{1}{2}(1+k)$ and $-\tfrac{1}{2}(1-k)$ are orthogonal projectors, this in particular implies
\be 
\mathfrak{shs}(K|L)[\lambda]\oplus \mathfrak{shs}(K|L)[1-\lambda] \subset \mathfrak{shs}_{K+L}[\lambda]\ .
\ee
Note that $\mathfrak{shs}(K|K)[\lambda] \cong \mathfrak{shs}_K[\lambda]$, as follows from the definition in (\ref{hu lambda definition}), and similarly for the other truncations. In fact, $K=L$ is the only case for which these algebras preserve some supersymmetry \cite{Vasiliev:1999ba}, see also \cite{Henneaux:2012ny,Candu:2014yva} for a related analysis. 

As is well known, see e.g.\ \cite[Section~7.2]{Gaberdiel:2012uj}, the $\mathfrak{shs}[\lambda]$ algebra contains the two bosonic higher spin algebras
\be
\mathfrak{shs}[\lambda]_{{\rm bos}} \cong \mathfrak{hs}[\lambda] \oplus \mathfrak{hs}[1-\lambda] \ . 
\ee
Similarly, we can consider the bosonic subalgebra of $\mathfrak{shs}(K|L)[\lambda]$, which is, by construction, generated by the block diagonal elements. Since the two block matrices along the diagonal, i.e.\ the terms proportional to $A$ and $D$ in (\ref{hu lambda definition}), do not mix, the bosonic subalgebra of $\mathfrak{shs}(K|L)[\lambda]$ is a direct sum of two bosonic subalgebras, which we may denote as 
\begin{align}
\mathfrak{shs}(K|L)[\lambda]_\text{bos}&\cong\mathfrak{hs}(K)[\lambda] \oplus \mathfrak{hs}(L)[1-\lambda]\ , \nonumber\\
\mathfrak{sho}(K|L)[\lambda]_\text{bos}&\cong\mathfrak{ho}(K)[\lambda] \oplus \mathfrak{ho}(L)[1-\lambda]\ , \nonumber\\
\mathfrak{shsp}(K|L)[\lambda]_\text{bos}&\cong\mathfrak{hsp}(K)[\lambda] \oplus \mathfrak{hsp}(L)[1-\lambda]\ . \label{bosonic subalgebras}
\end{align}

\section{CFT constructions}\label{sec:CFTconstructions}

It is straightforward to construct higher spin theories based on these higher spin algebras --- the construction of \cite{Prokushkin:1998bq,Prokushkin:1998vn} still goes through essentially unmodified --- and it is thus natural to ask what the corresponding CFT duals should be. Based on the general expectations from \cite{Campoleoni:2010zq,Gaberdiel:2011wb,Campoleoni:2011hg}, we can determine the relevant symmetry algebras via Drinfel'd-Sokolov reduction, and the analysis can be performed completely analogously to these constructions. The resulting ${\cal W}$ algebras then have the spin spectrum described in Table~\ref{tab:spin content}. We shall denote the ${\cal W}_\infty$ algebra associated to $\mathfrak{shs}(K|L)[\lambda]$ by $\mathfrak{s}\mathcal{W}_\infty(K|L)[\lambda]$.

Given the spin spectrum, one can turn the logic around and analyse the most general ${\cal W}$ algebra with that spin spectrum, imposing that the Jacobi identities have to be satisfied. We have done this analysis at low levels,\footnote{We have in particular considered $\mathfrak{so}\mathcal{W}_\infty(1|4)[c,\lambda]$, since this algebra will play an important role in Section~\ref{sec:applications}.} and it seems that each of these algebras is indeed characterized by just one parameter $\lambda$, in addition to the central charge $c$. We should mention that the corresponding analysis for the case $\mathfrak{s}\mathcal{W}_\infty(1|1)[\lambda] \cong \mathcal{W}_\infty^{\mathcal{N}=2}[\lambda]$ was already done in \cite{Candu:2012tr}, while that for $\mathfrak{s}\mathcal{W}_\infty(2|2)[\lambda] \cong \mathcal{W}_\infty^{\mathcal{N}=4}[\lambda]$ was done in \cite{Beccaria:2014jra}.

\subsection{The free field construction}\label{subsec:continuous orbifold}

The ${\cal N}=2$ superconformal ${\cal W}_\infty$ algebra, that agrees by construction with the algebra $\mathfrak{s}\mathcal{W}_\infty(1|1)[\lambda]$, has a free field realisation at $\lambda=0,1$, and one may therefore suspect that also $\mathfrak{s}\mathcal{W}_\infty(K|L)[\lambda]$ will have a free field realisation at $\lambda=0,1$. (In fact, it was this free field realisation that led us to consider this family of algebras in the first place.) 

In order to describe this free field realisation we consider $KN$ complex bosons and $LN$ complex fermions, where the bosons sit in the representation $(\mathbf{N},\bar{\mathbf{K}})\oplus (\bar{\mathbf{N}},\mathbf{K})$ of the algebra $\mathfrak{u}(K) \oplus \mathfrak{u}(N)$, and similarly for the fermions. (This is the natural generalisation of similar constructions that were considered before, see in particular \cite{Gaberdiel:2014cha, Gaberdiel:2014vca,Ferreira:2017zbh, Ferreira:2014xsr, Candu:2012ne}.) We then consider the $\mathfrak{u}(N)$-invariant combinations of these fields; as is familiar from these older constructions, see in particular \cite{Candu:2012jq}, the generators of the symmetry algebra consist entirely of the bilinears in the free fields. It is then straightforward to work out the spectrum of the generating fields, and one finds precisely the first column of Table~\ref{tab:spin content}, i.e.\ the algebra $\mathfrak{s}\mathcal{W}'_\infty(K|L)[\lambda]$. 
Actually, there is one small difference (which is why we added a prime to $\mathfrak{s}\mathcal{W}'_\infty$): at spin $1$, only the bilinears of the fermions give rise to currents, while the bilinears of the bosons only start at spin $s=2$. As a consequence, one actually obtains a slightly smaller algebra that misses the $\mathfrak{u}(K)$ algebra at $\lambda=0$ and the $\mathfrak{u}(L)$ algebra at $\lambda=1$. (Because of (\ref{duality}) the roles of the free bosons and fermions are interchanged as we replace $\lambda=0$ by $\lambda=1$.)\footnote{Note that this construction implies in particular that also the corresponding higher spin algebra $\mathfrak{shs}(K|L)[\lambda]$ contains these subalgebras at $\lambda=0,1$. We have checked explicitly that this is indeed compatible with the commutation relations of $\mathfrak{shs}(K|L)[\lambda]$.}  More details about this computation can be found in Appendix~\ref{app:details}.

We can also similarly perform the construction for the case of real (or pseudoreal) fields, where we think of the fields as transforming in the fundamental representation of 
$\mathfrak{so}(N)$ or $\mathfrak{sp}(N)$, respectively. This then leads to the second or third column of Table~\ref{tab:spin content}, except that again at spin $s=1$, the currents coming from the bilinears of the free bosons are missing. We denote the corresponding algebras as $\mathfrak{so}\mathcal{W}'_\infty(K|L)[\lambda]$ and 
$\mathfrak{sp}\mathcal{W}'_\infty(K|L)[\lambda]$, respectively.

%
%

\subsection{Coset constructions}

Given the by now familiar relation between free field realisations and coset theories, see e.g.\ \cite{Gaberdiel:2011aa,Gaberdiel:2014cha}, it is not difficult to guess the cosets that lead to these algebras. Indeed, the natural ansatz is\footnote{Note that here $k$ denotes the level, not to be confused with the Klein operator that appears in the definition of the oscillator algebra, see eq.~(\ref{defining commutators}).}
\begin{align} 
\frac{\mathfrak{su}(N+K)_k \oplus \mathfrak{u}(NL)_1}{\mathfrak{su}(N)_{k+L}\oplus \mathfrak{u}(1)_\kappa} \oplus \mathfrak{u}(1) \ . \label{u coset}
\end{align}
Here $\mathfrak{u}(NL)_1$ can be described in terms of $NL$ complex fermions in the representation $(\mathbf{N},\bar{\mathbf{L}}) \oplus (\bar{\mathbf{N}},\mathbf{L})$. The $\mathfrak{su}(N)_{k+L}$ algebra of the denominator is embedded into the numerator algebra as follows. The numerator algebra 
$\mathfrak{su}(N+K)_k$ contains the subalgebra $\mathfrak{su}(N)_k$ corresponding to a tracless blockmatrix of size $N$. The fermions generate furthermore $L$ $\mathfrak{su}(N)_1$ algebras, see e.g.\ \cite{DiFrancesco:1997nk}, and their diagonal sum is therefore a $\mathfrak{su}(N)_L$ algebra. Together with the subalgebra from the bosons this then describes the denominator $\mathfrak{su}(N)_{k+L}$ algebra. Finally, for the $\mathfrak{u}(1)_\kappa$ factor, we take the sum of the 
generator $\mathrm{diag}(K\mathds{1}_{N},-N\mathds{1}_{K })$ in $\mathfrak{su}(N+K)$ and the overall $\mathfrak{u}(1)$ current of the $\mathfrak{u}(NL)_1$ algebra (scaled by a factor of $(N+L)$ in order to match with standard conventions for $K=L=1$ and $K=L=2$); this leads to the total level of 
\be 
\kappa=k(N K^2+K N^2)+(N+L)^2 NL=k(N+K)NK+(N+L)^2NL\ .
\ee
The central $\mathfrak{u}(1)$ corresponds to the identity in the higher spin algebra. We may also consider the reduced coset without including this additional factor.

Notice that this coset includes a number of well-known examples as special cases. In particular, for $K=L=1$, the coset describes Kazama-Suzuki models with $\mathcal{N}=2$ superconformal symmetry, see e.g.\ \cite{Creutzig:2011fe,Candu:2012jq}. Another special case is $K=L=2$, where the model is known as a Wolf space coset with large ${\cal N}=4$ superconformal symmetry, see e.g.\ \cite{Gaberdiel:2013vva}.
\smallskip

In the limit of large level $k$, we obtain the free field theory of the previous section, except that the coset also contains a $\mathfrak{u}(K)_\infty$ algebra --- these are the generators of $\mathfrak{su}(N+K)$ that commute with the $\mathfrak{su}(N)$ subalgebra --- which has non-trivial commutation relations with the remaining bosons, but commutes with the fermions. This is precisely the $\mathfrak{u}(K)$ algebra that was missing in the free field realisation, see the discussion at the end of the previous subsection. It is also not difficult to check, using character arguments, see e.g.\ \cite{Candu:2012jq, Ferreira:2017zbh}, 
that the spin spectrum of the coset algebra agrees precisely with that of $\mathfrak{s}\mathcal{W}_\infty(K|L)[\lambda]$. 


A more interesting limit is given by the 't Hooft limit where  $N$ and $k$ are taken to infinity, while keeping the 't~Hooft parameter
\be 
\lambda\equiv\frac{N}{k+N} \label{t Hooft parameter}
\ee
fixed. The spin spectrum is independent of $\lambda$, and hence the resulting algebra must agree with the algebra $\mathfrak{s}\mathcal{W}_\infty(K|L)[\mu]$ for some choice of $\mu$ and $c$. An explicit calculation shows (see Appendix~\ref{app:thooft}) that in the above 't~Hooft limit $c\rightarrow \infty$ and $\mu=\lambda$, the 't Hooft parameter of \eqref{t Hooft parameter}.
Thus, these cosets provide a realisation of the $\mathfrak{s}\mathcal{W}_\infty(K|L)[\lambda]$ algebras.

From the coset perspective, the duality property \eqref{duality} can be motivated from level-rank duality, see e.g.\ \cite{Bouwknegt:1992wg}
\be \label{levelrank}
\frac{\mathfrak{su}(N)_m \oplus \mathfrak{su}(N)_n}{\mathfrak{su}(N)_{m+n}} \cong \frac{\mathfrak{su}(m+n)_N}{\mathfrak{su}(m)_N \oplus \mathfrak{su}(n)_N\oplus \mathfrak{u}(1)}\ . 
\ee
This slightly formal argument is spelled out in Appendix~\ref{app:LevelRank}.
\medskip

\noindent We can write down similar cosets for the orthogonal and symplectic analogs,
\begin{align}
\frac{\mathfrak{so}(N+K)_k \oplus \mathfrak{so}(NL)_1}{\mathfrak{so}(N)_{k+L}\oplus \mathfrak{u}(1)} \oplus \mathfrak{u}(1) \ ,  \label{socoset} \\[4pt]
\frac{\mathfrak{sp}(N+K)_k \oplus \mathfrak{sp
}(NL)_1}{\mathfrak{sp}(N)_{k+L}\oplus \mathfrak{u}(1)}\oplus \mathfrak{u}(1) \ .
\end{align}
A similar reasoning as above shows that their chiral algebras coincide, in the 't Hooft limit, with those of $\mathfrak{so}\mathcal{W}_\infty(K|L)[\lambda]$ and $\mathfrak{sp}\mathcal{W}_\infty(K|L)[\lambda]$, respectively. Here, $\lambda$ is again identified with the 't Hooft parameter \eqref{t Hooft parameter}.


\section{Applications to stringy large $\mathcal{N}=4$ holography} \label{sec:applications}

Our constructions were motivated by trying to understand the higher spin symmetry of string theory on 
$\mathrm{AdS}_3 \times \mathrm{S}^3 \times \mathrm{S}^3 \times \mathrm{S}^1$. This background was extensively studied in the literature \cite{Gukov:2004ym, Eberhardt:2017fsi, Elitzur:1998mm, Tong:2014yna, deBoer:1999gea,Baggio:2017kza}, and recently convincing evidence for the dual CFT was given in \cite{Eberhardt:2017pty}.
The background is characterized by three quantum numbers. $Q_1$ corresponds to the number of $\mathrm{D1}$-branes in the brane-construction, while $Q_5^+$ and $Q_5^-$ characterise the three-form flux through the two three-spheres, or equivalently the $\mathrm{D5}$-brane charges; for more details on the brane-construction, see  \cite{Gukov:2004ym, Eberhardt:2017pty}. It was claimed in \cite{Eberhardt:2017pty} that string theory with $Q_5^-/Q_5^+ \in \mathbb{Z}$ is dual to the symmetric orbifold theory
\be 
\mathrm{Sym}^{Q_1Q_5^+}(\mathfrak{su}(2)_{Q_5^-/Q_5^+-1} \oplus \mathfrak{u}(1) \oplus \mathfrak{so}(4)_1)\ ,\label{Sym Skappa}
\ee
where $\mathfrak{so}(4)_1$ describes four real fermions. In the following, we shall focus on the case $Q_5\equiv Q_5^+=Q_5^-$. Then the relevant dual CFT is
\be 
\mathrm{Sym}^{Q_1Q_5}(\mathfrak{u}(1) \oplus \mathfrak{so}(4)_1)\ .\label{Sym S0}
\ee
Note that $\mathfrak{u}(1) \oplus \mathfrak{so}(4)_1$ is the theory of one real boson and 4 real fermions, which is often denoted as $\mathcal{S}_0$ 
\cite{Eberhardt:2017fsi, Gukov:2004ym, Sevrin:1988ew}. This theory has large $\mathcal{N}=4$ supersymmetry. We will also write $N \equiv Q_1Q_5$.

The CFT dual of string theory on $\mathrm{AdS}_3 \times \mathrm{S}^3 \times \mathcal{M}_4$, where $\mathcal{M}_4= \mathbb{T}^4$ or $\mathrm{K3}$ lies on the same moduli space as the symmetric product orbifold of $\mathcal{M}_4$, see e.g.\ \cite{David:2002wn} for a review, and the same also seems to be true for certain orbifolds of ${\cal M}_4$ \cite{Datta:2017ert}. For the case of $\mathcal{M}_4= \mathbb{T}^4$ or $\mathrm{K3}$, the dual CFT contains an $\mathcal{N}=4$ higher spin algebra at least at certain points in the moduli space \cite{Gaberdiel:2014cha, Baggio:2015jxa}. By analogy, one might therefore expect that the symmetric product orbifold \eqref{Sym S0} should contain a natural $\mathcal{N}=4$ higher spin algebra. Somewhat surprisingly, this does not seem to be the case.

One reason why the situation for this background may be different is related to the form of the central charge of the dual CFT
\be 
c=\frac{6Q_1Q_5^+Q_5^-}{Q_5^++Q_5^-}\ .
\ee
Generically, this is not a half-integer, and thus the dual CFT is not a free theory. In particular, the usual bilinear construction of free fields that we employed above and that also appeared for the cases of $\mathcal{M}_4= \mathbb{T}^4$ and $\mathrm{K3}$ \cite{Gaberdiel:2014cha, Baggio:2015jxa} does no longer work: in general, bilinears of interacting fields do not close among themselves. This does not exclude the possibility that there could be a higher spin symmetry --- after all, in $2d$ interacting higher spin theories exist --- but  it has to arise by a more complicated mechanism. 

There are some exceptions to the above argument, namely when (here we assume without loss of generality as in \cite{Eberhardt:2017fsi} that $Q_5^- \ge Q_5^+$)
\be 
\gamma=\frac{Q_5^+}{Q_5^++Q_5^-} \in \tfrac{1}{12}\mathbb{Z}\ , \label{alpha}
\ee
which corresponds to particular ratios of the radii of the three-spheres in the geometry. These cases correspond to $\gamma=0$, $\tfrac{1}{12}$, $\tfrac{1}{6}$, $\tfrac{1}{4}$, $\tfrac{1}{3}$, $\tfrac{5}{12}$ and $\tfrac{1}{2}$. For $\gamma=0$, the large $\mathcal{N}=4$ algebra contracts to the small $\mathcal{N}=4$ algebra, and we obtain the same chiral algebra as for the symmetric orbifold of the $\mathbb{T}^4$ case; in particular, there is then a natural higher spin symmetry \cite{Gaberdiel:2014cha}. We will in the following mainly focus on the case of $\gamma=\tfrac{1}{2}$, for which the dual CFT is simplest \eqref{Sym S0}. We will also briefly discuss the other cases. As we shall see, there are natural higher spin algebras for all of these cases (except for $\gamma=\tfrac{5}{12}$ that is not even of the form (\ref{lambdaspec})), but none of them preserves the large ${\cal N}=4$ superconformal algebra.

\subsection{A $\mathbb{T}^2$-description}\label{T2algebra}
It was already noted in \cite{Eberhardt:2017pty}, that the chiral algebra of $\mathcal{S}_0$ can be identified with that of a supersymmetric torus $\mathbb{T}^2\cong \mathrm{S}^1 \times \mathrm{S}^1$. For this, we recall that the large $\mathcal{N}=4$ algebra has a natural $\mathcal{N}=2$ subalgebra \cite{Gukov:2004ym, Eberhardt:2017pty}. The $\mathfrak{u}(1)$ R-symmetry current of the $\mathcal{N}=2$ algebra is given by the sum of the two Cartan generators of the $\mathfrak{su}(2) \oplus \mathfrak{su}(2)$-algebra. In particular, w.r.t.~this $\mathcal{N}=2$ algebra, the fermions transform as
\be 
(\mathbf{2},\mathbf{2}) \longrightarrow (-\tfrac{1}{2}) \oplus 2 \cdot (0) \oplus (\tfrac{1}{2})\ .
\ee
Hence, there are two uncharged fermions, which we can bosonize, to obtain altogether two bosons and two fermions, i.e.~$\mathbb{T}^2$. It is then straightforward to check that the $\mathcal{N}=2$ algebra of the $\mathcal{S}_0$-theory becomes identified with the standard $\mathcal{N}=2$ algebra of $\mathbb{T}^2$. In particular, the generators of the $\mathcal{N}=2$ algebra in the $\mathbb{T}^2$-basis are bilinear in the fundamental fields. (In the $\mathcal{S}_0$-basis, however, the generators look quite complicated, as they involve trilinear terms in the fundamental fields.) Thus we conclude that the symmetric orbifold of ${\cal S}_0$ contains at least a $\mathfrak{s}\mathcal{W}_\infty(1|1)[0] \cong\mathcal{W}_\infty^{\mathcal{N}=2}[0]$ algebra \cite{Gaberdiel:2014vca}.

\subsection{The $\mathfrak{so}\mathcal{W}'_\infty(1|4)[0]$ algebra}\label{sow14}

There is however also another higher spin algebra contained in the symmetric orbifold of $\mathcal{S}_0$. In particular, we can consider the bilinear fields in the $N$ bosons and the $4N$ fermions, treating the $4N$ free fermions as real fermions that transform in the $4\cdot {\bf N}$ of $\mathfrak{so}(N)$.\footnote{For $N$ even, we could also take them to transform in the fundamental representation of $\mathfrak{sp}(N)$; then this will lead to the corresponding symplectic ${\cal W}_\infty$ algebra.} Using the arguments of Section~\ref{subsec:continuous orbifold}, this leads to the higher spin algebra $\mathfrak{so}\mathcal{W}'_\infty(1|4)[0]$. We should emphasize that it follows from our analysis that the corresponding higher spin algebra is $\mathfrak{sho}'(1|4)[0]$. In particular, one should therefore not expect that this algebra agrees with $\mathfrak{ho}(1|4)$ of \cite{Vasiliev:1999ba}; we have also shown this explicitly in Appendix~\ref{app:shoho}.


%

While this higher spin algebra is not supersymmetric, it is nonetheless attractive. First, it is the natural analogue of the small $\mathcal{N}=4$ construction of \cite{Gaberdiel:2014cha}. Second, the symmetry algebra is $\mathfrak{so}(4) \oplus \mathfrak{u}(1) \cong \mathfrak{su}(2) \oplus \mathfrak{su}(2)\oplus \mathfrak{u}(1)$,\footnote{The $\mathfrak{u}(1)$ arises from the linear term in the free boson which we may add.} so all isometries of the dual geometry are algebraically realized. In particular, the bosonic symmetry algebra of the system becomes
\be 
\mathfrak{u}(1) \oplus \mathfrak{so}\mathcal{W}_\infty(1)[0] \oplus \mathfrak{so}\mathcal{W}_\infty(4)[1]\ .
\ee
Finally, this symmetry algebra arises as the large level limit of a family of cosets, see eq.~(\ref{socoset}), and hence this higher spin subsector seems to arise from a natural higher spin--CFT duality. 

\subsection{Other values of $\gamma$}\label{sec:otheralpha}

For the other special values of $\gamma$ in \eqref{alpha} a similar analysis can be performed, using different free field realisations of $\mathfrak{su}(2)$, in particular (see  e.g.\ \cite{DiFrancesco:1997nk})
\begin{align}
\mathfrak{su}(2)_1 &\cong \mathfrak{u}(1)_1 \cong\text{boson at self-dual radius}\ , \\
\mathfrak{su}(2)_2 &\cong \mathfrak{so}(3)_1 \cong \text{three real fermions}\ , \\
\mathfrak{su}(2)_4 &\subset \mathfrak{su}(3)_1 \cong\text{two bosons on the $\mathfrak{su}(3)$-lattice}\ , \\
\mathfrak{su}(2)_{10} &\subset \mathfrak{sp}(4)_1 \cong \mathfrak{so}(5)_1 \cong\text{five real fermions}\ .
\end{align}
The first two cases are quantum equivalences, whereas the latter two are conformal embeddings. These exceptional cases cover $\gamma=\tfrac{1}{12}$, $\tfrac{1}{6}$, $\tfrac{1}{4}$ and $\tfrac{1}{3}$, but do not account for $\gamma=\tfrac{5}{12}$. Note that  $\gamma=\tfrac{5}{12}$ is also not covered by the proposal \eqref{Sym Skappa}, see eq.~(\ref{lambdaspec}),  and thus we shall not be able to say more about it.

Using these equivalences, it is straightforward to see that the following $\mathcal{W}_\infty$ algebras are subalgebras of the corresponding symmetric orbifolds
\begin{align}
\gamma=\tfrac{1}{3}&:\quad \mathfrak{so}\mathcal{W}_{\infty}'(2|4)[0]\ , \\
\gamma=\tfrac{1}{4}&:\quad \mathfrak{so}\mathcal{W}_{\infty}'(1|7)[0]\ , \\
\gamma=\tfrac{1}{6}&:\quad \mathfrak{so}\mathcal{W}_{\infty}'(3|4)[0]\ , \\
\gamma=\tfrac{1}{12}&:\quad \mathfrak{so}\mathcal{W}_{\infty}'(1|9)[0]\ .
\end{align}
Note that since we had to use conformal embeddings in the last two cases, we need an off-diagonal modular invariant to embed these $\mathcal{W}_\infty$ algebras into the symmetric orbifold. We should also mention that for $\gamma=\frac{1}{3}$ and $\gamma=\frac{1}{6}$, there are two further constructions that are possible provided that the compactification radius of the free boson in the $\mathcal{S}_0$-theory takes a special value, so that 
\begin{align}
\mathfrak{u}(2)_1 &\cong \text{two complex fermions} \cong \text{four real fermions}\ , \\
\mathfrak{u}(2)_4 &\cong \mathfrak{u}(3)_1 \cong \text{three complex fermions} \cong \text{six real fermions}\ .
\end{align}
For these special radii, we could thus also construct an $\mathfrak{so}\mathcal{W}_{\infty}'(0|8)[0]$ algebra for $\gamma=\tfrac{1}{3}$, and an $\mathfrak{so}\mathcal{W}_{\infty}'(0|10)[0]$ algebra for $\gamma=\tfrac{1}{6}$. As before for the case $\gamma=\frac{1}{2}$, we could have also considered the corresponding symplectic version, see footnote~6. 
Finally, we note that the small $\mathcal{N}=4$ case of $\gamma=0$ fits naturally into this series since we have
\be 
\mathfrak{su}(2)_\infty \cong \mathfrak{u}(1)_\infty^3 \cong\text{three non-compact bosons}\ ,
\ee
and hence
\be 
\gamma=0:\quad \mathfrak{so}\mathcal{W}_{\infty}'(4|4)[0] \supset \mathfrak{su}\mathcal{W}_{\infty}'(2|2)[0] \cong \mathcal{W}^{\mathcal{N}=4}_\infty[0]\ .
\ee
Incidentally, given the results of \cite{Ferreira:2017pgt}, it is quite natural that only the even spin subalgebra of the higher spin square appears.

\section{Conclusions}\label{sec:conclusions}

In this paper we have constructed a new family of higher spin algebras; they arise upon restricting a suitable matrix extension of the familiar $\mathfrak{shs}[\lambda]$ algebra, see eq.~(\ref{hu lambda definition}). We have identified CFT coset duals realising these symmetry algebras, and thus natural higher spin--CFT dual pairs. Our construction was motivated by trying to identify higher spin subsectors in the CFT dual of string theory on AdS$_3\times {\rm S}^3 \times {\rm S}^3 \times {\rm S}^1$. Indeed, the higher spin algebras we have constructed appear naturally for some special values of $\gamma$, i.e.\ special ratios of the sizes of the two ${\rm S}^3$'s. 

Generically, the higher spin algebras we have constructed are non-supersymmetric, and this is indeed the case for the examples that arise in the context of AdS$_3\times {\rm S}^3 \times {\rm S}^3 \times {\rm S}^1$. It is not clear to us what the fundamental reason for this is. Note that, for $\gamma=\frac{1}{2}$, there exists also a supersymmetric higher spin subsector, as already identified in \cite{Eberhardt:2017pty}. In particular, our analysis does not preclude that there are also other (supersymmetric) higher spin algebras one may find in these theories.

\section*{Acknowledgements} This paper is partially based on the Master thesis of one of us (IR). We thank Kevin Ferreira and Wei Li for useful conversations about related topics. The research of LE and MRG is supported in parts  by the NCCR SwissMAP, funded by the Swiss National Science Foundation. LE gratefully acknowledges the hospitality of the Institute for Advanced Study during the later stages of this work. MRG is grateful to the BICMR at Beijing University for hospitality during the final stages of this work.

\appendix

\section{Orthogonal and symplectic algebras}\label{sec:orthsymp}

Let us recall a few important representations of $\mathfrak{so}(M)$ and $\mathfrak{sp}(M)$.\footnote{In the symplectic case we always assume that $M$ is even; the fundamental representation of $\mathfrak{sp}(M)$ has then dimension $M$.} We shall need the adjoint representation, the fundamental representation and the symmetric representation of $\mathfrak{so}(M)$. We shall denote them by their Dynkin labels $[0,1,0,\dots]$, $[1,0,\dots]$ and $[2,0,\dots]\oplus [0,0,\dots]$,\footnote{The symmetric representation is not irreducible, but we shall always only need the combination $[2,0,\dots]\oplus [0,0,\dots]$.} respectively. We have
\begin{align} 
&\mathrm{dim}([1,0,\dots])=M\ , \quad \mathrm{dim}\left([0,1,0,\dots]\right)=\tfrac{1}{2}M(M-1)\nonumber\\
&\qquad\text{and}\quad\mathrm{dim}([2,0,\dots]\oplus [0,0,\dots])=\tfrac{1}{2}M(M+1)\ .
\end{align}
Similar considerations apply to $\mathfrak{sp}(M)$, where we have
\begin{align} 
&\mathrm{dim}([1,0,\dots])=M\ , \quad \mathrm{dim}\left([2,0,\dots]\right)=\tfrac{1}{2}M(M+1)\nonumber\\
&\qquad\text{and}\quad \mathrm{dim}([0,1,0\dots]\oplus [0,0,\dots])=\tfrac{1}{2}M(M-1)\ .
\end{align}
On $\mathbb{R}^{K+L}$, we are using the symplectic form $\Omega=\Omega_K \oplus \Omega_L$, where $\Omega_K$ and $\Omega_L$ are symplectic forms on $\mathbb{R}^K$ and $\mathbb{R}^L$, respectively. (Recall that we are always assuming that $K$ and $L$ even in the symplectic case.)

\section{Details of the free field calculation} \label{app:details}
Here we present some more details about the free field realisation of the $\mathfrak{s}{\cal W}_\infty(K|L)[\lambda]$ algebra at $\lambda=0$. We consider $NK$ free complex bosons $ \partial \phi^{\bar{\imath},a}, \partial \bar{\phi}^{i,\bar{a}}$ and $NL$ free complex fermions $\psi^{\bar{\imath},\alpha}, \bar{\psi}^{i,\bar{\alpha}}$, where 
$i,\bar{\imath}=1,\dots,N$, $a,\bar{a}=1,\dots,K$, and $\alpha,\bar{\alpha}=1,\dots,L$. We define quasi-primary fields, following \cite{Bergshoeff:1990}, as
\begin{align}
W^{\alpha \bar{\beta}}_{\mathrm{F},h}(z) &= n_{W_{\mathrm{F},h}}\sum_{k=0}^{h-1} \sum_{i,\, \bar{\imath}=1}^N\delta_{i,\bar{\imath}}(-1)^k \binom{h-1}{k}^2\!:\!\partial^k\psi^{\bar{\imath},\alpha}\partial^{h-k-1}\bar{\psi}^{i,\bar{\beta}}\!:\!(z) \ ,\\
W^{a \bar{b}}_{\mathrm{B},h}(z) &= n_{W_{\mathrm{B},h}}\sum_{k=0}^{h-2}\sum_{i,\, \bar{\imath}=1}^N \delta_{i,\bar{\imath}}\frac{(-1)^k}{h-1} \binom{h-1}{k}\binom{h-1}{k+1}\!:\!\partial^{k+1}\phi^{\bar{\imath},a}\partial^{h-k-1}\bar{\phi}^{i,\bar{b}}\!:\!(z)\ ,\\
Q^{a\bar{\alpha}}_h(z) &= n_{Q_h} \sum_{k=0}^{h-\frac{3}{2}}\sum_{i,\, \bar{\imath}=1}^N \delta_{i,\bar{\imath}}(-1)^k \binom{h-\frac{3}{2}}{k}\binom{h-\frac{1}{2}}{k}\!:\!\partial^{h-k-\frac{1}{2}}\phi^{\bar{\imath},a}\partial^k\bar{\psi}^{i,\bar{\alpha}}\!:\!(z)\ ,\\
\bar{Q}^{\bar{a}\alpha}_h(z) &= n_{\bar{Q}_h} \sum_{k=0}^{h-\frac{3}{2}} \sum_{i,\, \bar{\imath}=1}^N\delta_{i,\bar{\imath}}(-1)^{k} \binom{h-\frac{3}{2}}{k}\binom{h-\frac{1}{2}}{k}\!:\!\partial^{h-k-\frac{1}{2}}\bar{\phi}^{i,\bar{a}} \partial^k\psi^{\bar{\imath},\alpha}\!:\!(z)\ ,
\end{align}
where we sum over repeated indices of $i$, $\bar{\imath}$, and the $n_{*}$ are some arbitrary normalisation constants. $W^{a \bar{b}}_{B,h}$ and $W^{\alpha \bar{\beta}}_{F,h}$ transform in the adjoint representation of the $\mathfrak{u}(K)$ or the $\mathfrak{u}(L)$ algebra, respectively. $Q^{a\bar{\alpha}}_h$ and $\bar{Q}^{\bar{a}\alpha}_h$ have half-integer spin and transform in the bifundamental representation $(\mathbf{K},\mathbf{\bar{L}})$ or $(\mathbf{\bar{K}},\mathbf{L})$ of the direct sum of the above algebras. In this basis the stress-energy tensor is given by the sum of the traces
\be
	T(z) =  T_{\mathrm{B}}(z) + T_{\mathrm{F}}(z) = \frac{1}{n_{W_{\mathrm{B},2}}} \sum_{a,\, \bar{a}=1}^K\delta_{a,\bar{a}}W^{a \bar{a}}_{\mathrm{B},2}(z) - \frac{1}{2 n_{W_{\mathrm{F},2}}}\sum_{\alpha,\, \bar{\alpha}=1}^L \delta_{\alpha,\bar{\alpha}} W^{\alpha \bar{\alpha}}_{\mathrm{F},2}(z)\ .
\ee
We have implemented the fields up to spin 4 and for different combinations of small $K,L$ explicitly. 
We know, on general grounds, see in particular \cite{Blumenhagen:1990jv}, that the commutators must take the form 
\begin{align}
[(W^{\alpha \bar{\beta}}_{\mathrm{F},h_1})_m,(W^{\gamma \bar{\delta}}_{\mathrm{F},h_2})_n] &= \sum_{\substack{h\in \mathbb{Z}_+ \\  h < h_1+h_2}} p_{\mathrm{F}}^{h_1h_2h}(m,n) \, \Big( \delta^{\gamma \bar{\beta}} \, W^{\alpha \bar{\delta}}_{\mathrm{F},h} + (-1)^{h_1+h_2+h} \delta^{\alpha \bar{\delta}} \, W^{\gamma \bar{\beta}}_{\mathrm{F},h}\Big)_{m+n}\ , \nonumber\\
[(W^{a \bar{b}}_{\mathrm{B},h_1})_m,(W^{c \bar{d}}_{\mathrm{B},h_2})_n] &= \!\!\sum_{\substack{h\in \mathbb{Z}_+ \\  1 < h < h_1+h_2}} \!\! p_{\mathrm{B}}^{h_1h_2h}(m,n) \,\Big( \delta^{c \bar{b}} \, W^{a \bar{d}}_{\mathrm{B},h} + (-1)^{h_1+h_2+h} \delta^{a \bar{d}} \, W^{c \bar{b}}_{\mathrm{B},h}\Big)_{m+n}\ ,\nonumber \\
[(W^{\alpha \bar{\beta}}_{\mathrm{F},h_1})_m,(Q^{a \bar{\gamma}}_{h_2})_r] &= \sum_{\substack{h\in (\mathbb{Z}_++\frac{1}{2}) \\  h < h_1+h_2}} q_{\mathrm{F}}^{h_1h_2h}(m,r) \, \delta^{\alpha \bar{\gamma}} \, (Q^{a \bar{\beta}}_{h})_{m+r}\ ,\label{WF Q}\\
[(W^{a \bar{b}}_{\mathrm{B},h_1})_m,(Q^{c \bar{\alpha}}_{h_2})_r] &= \sum_{\substack{h\in (\mathbb{Z}_++\frac{1}{2}) \\  h < h_1+h_2}} q_{\mathrm{B}}^{h_1h_2h}(m,r) \, \delta^{c \bar{\beta}} \, (Q^{a \bar{\alpha}}_{h})_{m+r}\ ,\label{WB Q}\\
\{(Q^{a \bar{\alpha}}_{h_1})_r,(\bar{Q}^{\bar{b} {\beta}}_{h_2})_s\} &=\sum_{\substack{h\in \mathbb{Z}_+ \\ 1< h < h_1+h_2}} \Big( o_{\mathrm{F}}^{h_1h_2h}(r,s) \ \delta^{a \bar{b}} \, W^{\alpha \bar{\beta}}_{\mathrm{F},h}  + o_{\mathrm{B}}^{h_1h_2h}(r,s) \, \delta^{\alpha \bar{\beta}} \, W^{a \bar{b}}_{\mathrm{B},h}\Big)_{r+s}\nonumber\\
&\qquad+ o_{\mathrm{F}}^{h_1h_21}(r,s) \ \delta^{a \bar{b}} \ (W^{\alpha \bar{\beta}}_{\mathrm{F},1})_{r+s} \ ,\label{Q Q}
\end{align}
where the functions $p_F^{h_1h_2h},p_B^{h_1h_2h},\dots,o_B^{h_1h_2h}$ are polynomials in the mode numbers $m,n \in \mathbb{Z}$ and $r,s \in \mathbb{Z}+\frac{1}{2}$, with coefficients that depend on the conformal weights $h_1,h_2,h$, see e.g.\ \cite{Blumenhagen:1990jv}. Furthermore, their dependence on the parameters $a, \bar{a}$ and $\alpha,\bar{\alpha}$ is entirely fixed by the representation theory, see Table~\ref{tab:spin content}. 
With the help of  the \texttt{Mathematica} package of Thielemans \cite{Thielemans:1991} we have confirmed that the structure constants stablilize indeed for $K\ge 2$ and $L \ge 2$; suppressing the group theoretic index structure (and setting all normalisations $n_*=1$), we have found explicitly
\begin{alignat*}{2}
&{[W_{\mathrm{F},1},W_{\mathrm{F},1}]}:&& \quad p_{\mathrm{F}}^{111} = 1\ ,\\[5pt]
&{[W_{{\mathrm{F}},1},Q_{\frac{3}{2}}]}:&&\quad q_{\mathrm{F}}^{1\frac{3}{2}\frac{3}{2}} = -1\ ,\\[5pt]
&{\{Q_{\frac{3}{2}},\bar{Q}_{\frac{3}{2}}\}}:&& \quad o_{\mathrm{F}}^{\frac{3}{2}\frac{3}{2}1} = - \frac{1}{2}\left(r-s\right)\ , \quad o_{\mathrm{F}}^{\frac{3}{2}\frac{3}{2}2} = - \frac{1}{2}\ , \quad o_{\mathrm{B}}^{\frac{3}{2}\frac{3}{2}2} =  1\ ,\\[5pt]
&{[W_{{\mathrm{F}},1},W_{{\mathrm{F}},2}]}:&& \quad p_{\mathrm{F}}^{121}=- m\ , \quad p_{\mathrm{F}}^{122}=1\ ,\\[5pt]
&{[W_{{\mathrm{F}},2},Q_{\frac{3}{2}}]}:&& \quad q_{\mathrm{F}}^{2\frac{3}{2}\frac{3}{2}}= \frac{1}{3}\left(2r-m\right)\ , \quad q_{\mathrm{F}}^{2\frac{3}{2}\frac{5}{2}}= \frac{2}{3}\ , \\[5pt]
&{[W_{{\mathrm{B}},2},Q_{\frac{3}{2}}]}:&& \quad q_{\mathrm{B}}^{2\frac{3}{2}\frac{3}{2}}= -\frac{1}{3}\left(2r-m\right)\ , \quad q_{\mathrm{B}}^{2\frac{3}{2}\frac{5}{2}}= \frac{1}{3}\ , \\[5pt]
&{[W_{{\mathrm{F}},2},W_{{\mathrm{F}},2}]}:&& \quad p_{\mathrm{F}}^{221} = \frac{1}{3}\left(m^2-mn+n^2-1\right)\ , \quad p_{\mathrm{F}}^{222} = -\left(m-n\right) \ ,\quad p_{\mathrm{F}}^{223} =  \frac{2}{3}\ ,\\[5pt]
&{[W_{{\mathrm{B}},2},W_{{\mathrm{B}},2}]}:&& \quad p_{\mathrm{B}}^{222} = \frac{1}{2}\left(m-n\right)\ , \quad p_{\mathrm{B}}^{223} = - \frac{1}{2}\ ,\\[7pt]
&{[W_{{\mathrm{F}},1},Q_{\frac{5}{2}}]}:&&\quad q_{\mathrm{F}}^{1\frac{5}{2}\frac{3}{2}} = 2 m\ , \quad q_{\mathrm{F}}^{1\frac{5}{2}\frac{5}{2}} = -1\ ,\\[7pt]
&{\{Q_{\frac{3}{2}},\bar{Q}_{\frac{5}{2}}\}}:&& \quad o_{\mathrm{F}}^{\frac{3}{2}\frac{5}{2}1} = -\left(r^2-\frac{1}{4}\right)\ , \quad o_{\mathrm{F}}^{\frac{3}{2}\frac{5}{2}2} = -\frac{1}{2} \left(3r-s\right)\ ,\\
& &&\quad o_{\mathrm{B}}^{\frac{3}{2}\frac{5}{2}2} = -\frac{1}{2} \left(3r-s\right)\ , \quad o_{\mathrm{F}}^{\frac{3}{2}\frac{5}{2}3}= -\frac{1}{2}\ , \quad o_{\mathrm{B}}^{\frac{3}{2}\frac{5}{2}3}= \frac{3}{2}\ ,\\[5pt]
&{[W_{{\mathrm{F}},2},Q_{\frac{5}{2}}]}:&& \quad q_{\mathrm{F}}^{2\frac{5}{2}\frac{3}{2}} = \left(m^2-\frac{2}{3}mr+\frac{1}{3}r^2-\frac{3}{4}\right)\ , \quad q_{\mathrm{F}}^{2\frac{5}{2}\frac{5}{2}} = -\frac{14}{5}\left(\frac{m}{2}-\frac{r}{3}\right)\ , \\
& && \quad q_{\mathrm{F}}^{2\frac{5}{2}\frac{7}{2}} = \frac{3}{5}\left(\frac{m}{2}-\frac{r}{3}\right)\ ,\\[5pt]
&{[W_{{\mathrm{B}},2},Q_{\frac{5}{2}}]}:&& \quad q_{\mathrm{B}}^{2\frac{5}{2}\frac{3}{2}} = \frac{1}{2}\left(m^2-\frac{2}{3}mr+\frac{1}{3}r^2-\frac{3}{4}\right)\ , \quad q_{\mathrm{B}}^{2\frac{5}{2}\frac{5}{2}} = \frac{8}{5}\left(\frac{m}{2}-\frac{r}{3}\right)\ , \\
& &&\quad q_{\mathrm{B}}^{2\frac{5}{2}\frac{7}{2}} = \frac{3}{10}\left(\frac{m}{2}-\frac{r}{3}\right)\ ,\\[5pt]
&{\{Q_{\frac{5}{2}},\bar{Q}_{\frac{5}{2}}\}}:&& \quad o_{\mathrm{F}}^{\frac{5}{2}\frac{5}{2}1} = \frac{1}{4}\left(r^3-r^2s-\frac{5}{2}\left(r-s\right)+rs^2-s^3\right)\ ,\\
& && \quad o_{\mathrm{F}}^{\frac{5}{2}\frac{5}{2}2} = \frac{21}{20}\left(r^2-\frac{4}{3}rs+s^2-\frac{3}{2}\right)\ ,\\
& && \quad o_{\mathrm{B}}^{\frac{5}{2}\frac{5}{2}2} = -\frac{6}{5}\left(r^2-\frac{4}{3}rs+s^2-\frac{3}{2}\right)\ ,\quad o_{\mathrm{F}}^{\frac{5}{2}\frac{5}{2}3} = \frac{5}{4}\left(r-s\right)\ , \\
& && \quad o_{\mathrm{B}}^{\frac{5}{2}\frac{5}{2}3} = 3\left(r-s\right)\ , \quad o_{\mathrm{F}}^{\frac{5}{2}\frac{5}{2}4} = -\frac{9}{5}\ , \quad o_{\mathrm{B}}^{\frac{5}{2}\frac{5}{2}4} = \frac{9}{20}\ .
\end{alignat*} 
We have checked that these structure constants coincide exactly with those of the higher spin algebra $\mathfrak{shs}(K|L)[0]$, provided one chooses compatible normalisations for the higher spin generators. For this we have implemented the oscillator relations \eqref{defining commutators} in \texttt{Mathematica}, and then defined the generators of $\mathfrak{shs}(K|L)[\lambda]$ using \eqref{hu lambda definition}. We have confirmed explicitly that for $\lambda = 0$ the spin $s=1$ $\mathfrak{u}(K)$ generators decouple (see \eqref{nu dep1} for the lowest-dimensional occurrence of this phenomenon), and that the remaining structure constants agree with the results shown above.


\section{'t~Hooft limit of the coset construction}\label{app:thooft}

In this appendix we sketch how we have calculated the $\lambda$ parameter from the coset viewpoint. Let us denote the $\mathfrak{su}(N+K)$ currents of the numerator by $J^{u\bar{v}}$, where $u,v=1,\dots,N+K$,\footnote{We are a bit cavalier here with the external $\mathfrak{u}(1)$ factors since they do not affect the structure constants in the 't~Hooft limit.} 
and the corresponding fermions by $\psi^{\bar{\imath},\alpha}$ (along with their complex conjugates), where $i,\bar{\imath}=1,\dots,N$ and $\alpha,\bar{\alpha}=1,\dots,L$. Then the 
spin $1$ affine algebras are given by 
\be 
W_{1,\mathrm{B}}^{a\bar{b}}=J^{n+a,n+\bar{b}}(z)\ , \quad \hbox{and} \quad W_{1,\mathrm{F}}^{\alpha\bar{\beta}}
=\sum_{i,\bar{\imath}} \delta_{i,\bar{\imath}}(\psi^{\bar{\imath},\alpha}\bar{\psi}^{i,\bar{\beta}})(z)\ ,
\ee
and have level $k$ and $N$, respectively. For the spin $\tfrac{3}{2}$ fields we have (up to an arbitrary normalization) 
\be 
Q_\frac{3}{2}^{a \bar{\alpha}}(z)=n_{Q_{\frac{3}{2}}}\sum_{i,\bar{\imath}} \delta_{i,\bar{\imath}}(J^{n+a,\bar{\imath}}\bar{\psi}^{i,\bar{\alpha}})(z)\ , \quad \bar{Q}_\frac{3}{2}^{\bar{a} \alpha}(z)=n_{\bar{Q}_{\frac{3}{2}}}\sum_{i,\bar{\imath}} \delta_{i,\bar{\imath}}(J^{i,n+\bar{a}}\psi^{\bar{\imath},\alpha})(z)\ .
\ee
Their OPE is straightforward to calculate, and we find 
\begin{align}
Q_\frac{3}{2}^{a \bar{\alpha}}(z)\bar{Q}_\frac{3}{2}^{\bar{b} \beta}(w) &\sim n_{Q_{\frac{3}{2}}} n_{\bar{Q}_{\frac{3}{2}}}\frac{N \delta^{\bar{\alpha},\beta}W_{1,\mathrm{B}}^{a \bar{b}}(w)+k \delta^{a,\bar{b}}W_{1,\mathrm{F}}^{\bar{\alpha} \beta}(w)  }{(z-w)^2}\nonumber\\
&\qquad\qquad\qquad+(\mathfrak{u}(1)\text{-currents}+ \text{first order poles)}.
\end{align}
We can extract the structure constants for the corresponding modes, and in the notation of Appendix~\ref{app:details}, we find 
\be 
o^{\frac{3}{2}\frac{3}{2}1}_{\mathrm{F}}(r,s)=-\frac{1}{2}n_{Q_{\frac{3}{2}}} n_{\bar{Q}_{\frac{3}{2}}}k(r-s)\ , \quad o^{\frac{3}{2}\frac{3}{2}1}_{\mathrm{B}}(r,s)=-\frac{1}{2}n_{Q_{\frac{3}{2}}} n_{\bar{Q}_{\frac{3}{2}}}N(r-s)\ .
\ee
This is to be compared with the structure constants of the higher spin algebra, for which we find 
\be 
o^{\frac{3}{2}\frac{3}{2}1}_{\mathrm{F}}(r,s)=-\frac{1}{2}(1-\lambda)(r-s)\ , \quad o^{\frac{3}{2}\frac{3}{2}1}_{\mathrm{B}}(r,s)=-\frac{1}{2}\lambda (r-s)\ .
\ee
Matching the two expressions gives then \eqref{t Hooft parameter}.

\section{Level Rank Duality}\label{app:LevelRank}

The duality property \eqref{duality} can be motivated by the following set of transformations
\begin{align}
\frac{\mathfrak{su}(N+K)_k \oplus \mathfrak{u}(NL)_1}{\mathfrak{su}(N)_{k+L}\oplus\mathfrak{u}(1)} &\cong \frac{\mathfrak{su}(N+K)_k \oplus \mathfrak{su}(N)_L\oplus \mathfrak{su}(L)_N\oplus \mathfrak{u}(1)}{\mathfrak{u}(N)_{k+L}\oplus \mathfrak{u}(1)}\\
&\cong \frac{\mathfrak{su}(N+K)_k}{\mathfrak{su}(N)_k\oplus \mathfrak{su}(K)_k\oplus\mathfrak{u}(1)}\oplus \frac{\mathfrak{su}(N)_k \oplus \mathfrak{su}(N)_L}{\mathfrak{su}(N)_{k+L}}\nonumber\\
&\qquad\oplus \mathfrak{su}(K)_k \oplus \mathfrak{su}(L)_N \oplus \mathfrak{u}(1)\\
&\cong \frac{\mathfrak{su}(k)_N \oplus \mathfrak{su}(k)_K}{\mathfrak{su}(k)_{N+K}}\oplus \frac{\mathfrak{su}(k+L)_N}{\mathfrak{su}(k)_N\oplus \mathfrak{su}(L)_N \oplus \mathfrak{u}(1)}\nonumber\\
&\qquad\oplus \mathfrak{su}(K)_k \oplus \mathfrak{su}(L)_N \oplus \mathfrak{u}(1)\\
&\cong \frac{\mathfrak{su}(k+L)_N\oplus \mathfrak{su}(K)_k\oplus \mathfrak{su}(k)_K\oplus \mathfrak{u}(1)}{\mathfrak{su}(k)_{N+K} \oplus \mathfrak{u}(1)} \\
&\cong \frac{\mathfrak{su}(k+L)_N \oplus \mathfrak{u}(kK)_1}{\mathfrak{su}(k)_{N+K}\oplus\mathfrak{u}(1)} \ .
\end{align}
Here, we have used the standard level-rank duality (\ref{levelrank}), as well as the conformal embeddings 
\be 
\mathfrak{su}(M)_N \oplus \mathfrak{su}(N)_M \subset \mathfrak{su}(MN)_1\ .
\ee
We should stress, however,  that these arguments should be taken with a grain of salt since splitting up cosets and recombining them (as well as using conformal embeddings) are not actual isomorphisms. 

\section{The $\mathfrak{sho}'(1|4)[\lambda]$ and $\mathfrak{ho}(1|4)$ algebra}\label{app:shoho}
In this section we want to show explicitly why the $\mathfrak{so}\mathcal{W}'_\infty(1|4)[0]$ algebra of Section~\ref{sow14} does not correspond to the $\mathfrak{ho}(1|4)$ algebra of \cite{Vasiliev:1999ba}, but to $\mathfrak{sho}'(1|4)[0]$. This was the original motivation for our construction. 

As already explained in Section~\ref{matrix extension},  $\mathfrak{sho}'(1|4)[\lambda]$ is isomorphic to $\mathfrak{ho}(1|4)$ at $\lambda = \frac{1}{2}$. (Strictly speaking, the argument was given for the unitary version of the algebra, but the orthogonal case works similarly.)
It remains to show that $\mathfrak{sho}'(1|4)[\frac{1}{2}] \ncong \mathfrak{sho}'(1|4)[0]$.  As the truncation to the orthogonal subalgebra does not affect the  dependence on $\nu = 2 \lambda -1$, it suffices to investigate the structure constants of $\mathfrak{shs}(K|L)[\lambda]$. The normalization of the spin 1 fields is determined by requiring that they generate $\mathfrak{u}(K)\oplus \mathfrak{u}(L)$. The spin $\tfrac{3}{2}$ commutation relations are schematically\footnote{To keep notation light, we suppress the mode numbers and group-theoretical indices.}
\be 
	\{Q_{\frac{3}{2}},\bar{Q}_\frac{3}{2}\} =\lambda\, W_{\mathrm{B},1} + (1-\lambda) \, W_{\mathrm{F},1} + \text{spin-2 fields} \ , \label{nu dep1} 
\ee
which, in particular, fixes the normalizations of the spin $\tfrac{3}{2}$ fields. It is then clear that the algebras for different values of $\lambda$ are not isomorphic, except for the isomorphism that interchanges $\lambda$ with $1-\lambda$ and simultaneously $W_{\mathrm{B},1}$ with $W_{\mathrm{F},1}$ --- this is the duality that was discussed already around eq.~\eqref{duality}.

\bibliographystyle{JHEP}

\end{document}